\documentclass[12pt]{article}
\pdfoutput=1
\usepackage{hyperref}
\usepackage{color}
\usepackage{graphicx}
\usepackage{amsmath}
\usepackage{amssymb}
\usepackage{xspace}
\usepackage{soul}

\makeatletter
\@addtoreset{equation}{section}

\makeatletter
\renewcommand\section{\@startsection {section}{1}{\z@}%
                                   {-3.5ex \@plus -1ex \@minus -.2ex}
                                   {2.3ex \@plus.2ex}%
                                   {\normalfont\large\bfseries}}
\renewcommand\subsection{\@startsection{subsection}{2}{\z@}%
                                     {-3.25ex\@plus -1ex \@minus -.2ex}%
                                     {1.5ex \@plus .2ex}%
                                     {\normalfont\bfseries}}

\newcommand{\be}{\begin{equation}}
\newcommand{\ee}{\end{equation}}
\newcommand{\beq}{\begin{eqnarray}}
\newcommand{\eeq}{\end{eqnarray}}

\newcommand{\gone}[1]{{}}

\begin{document}
\begin{titlepage}

\rule{0ex}{0ex}

\vfil

\begin{center}

{\bf \Large
Comments on   the negative specific heat of the $T \bar T$  deformed symmetric product CFT
}

\vfil

Soumangsu Chakraborty$^1$,  Akikazu Hashimoto$^2$

\vfil

{}$^1$ Universit\'e Paris-Saclay, CNRS, CEA, \\Institut de Physique Th\'eorique, 91191 Gif-sur-Yvette, France

{}$^2$ Department of Physics, University of Wisconsin, Madison, WI 53706, USA

\vfil

\end{center}

\begin{abstract}
\noindent We show that the $T \bar T$ deformation of conformal field theories whose entropy grows as $S(E) \sim E^\gamma$ for $\gamma > 1/2$ exhibits negative specific heat in its microcanonical thermodynamic function  $S({\cal E})$. We analyze the large  $N$ symmetric product CFT as a concrete example of a CFT with this property and compute the thermodynamic functions such as $S({\cal E})$ and ${\cal E}(T)$. The negative specific heat in the microcanonical data is interpreted as signaling the first order phase transition when the system is coupled to a heat bath. 
\end{abstract}
\vspace{0.5in}

\end{titlepage}

\section{Introduction}

Recently, there have been some interest in the subject of $\mu T \bar T$ deformed quantum field theories \cite{Smirnov:2016lqw,Cavaglia:2016oda}. This deformation was defined originally for quantum field theories in 1+1 dimensions, for which the deforming operator is $\mu \det T = \mu (T_{z z} \bar T_{\bar z \bar z}- T_{z \bar z}^2)$. This deformation is special in that the energy spectrum of the deformed theory on cylindrical space-time $\mathbb{R} \times S^1$, of radius $R$, can be deduced from the spectrum of the undeformed theory via a flow equation of the form
\be
{\partial \over \partial \mu} {\cal E}_i(R,\mu)  + {\cal E}_i(R,\mu) {\partial \over \partial R} {\cal E}_i(R,\mu) + {P_i^2(R) \over R} = 0 \ee
where $i$ is the index variable labeling the states, and 
\be {\cal E}_i(R,\mu=0) = E_i(R), \qquad P_i(r) = {n_i \over R} \ee
is the spectrum of the undeformed theory. This flow equation can be recognized as the Burgers equation and admits a formal solution which can be found for instance in (5.5) of \cite{Smirnov:2016lqw}. If we restrict to the case that the undeformed theory is conformal, we can impose that
\be R {\cal E}_i(R,\mu) \ee
is independent of $R$, and the deformed spectrum can be written more explicitly as
\be {\cal E}_i = \sqrt{{R^2 \over 4 \mu^2} +{R E_i \over \mu} + P_i^2} - {R \over 2 \mu} \ .  \label{def} \ee
There is some qualitative differences in the nature of the deformed spectrum depending on the sign of $\mu$ \cite{McGough:2016lol}. We will primarily focus on the case where $\mu>0$ so that for every state labeled by $i$ of the undeformed spectrum, there is a deformed state with real energy ${\cal E}_i$. 

Following common practice, we will loosely  refer to the $\mu \det T$ deformation as the $\mu T \bar T$ deformation in this note. What is remarkable about this deformation is the fact that the theory is well defined as a quantum theory, in the the sense that the spectrum is well defined, despite the fact that the deforming operator is irrelevant. This is contrary to the expectation based on conventional wisdom in renormalization group theory. The point of \cite{Smirnov:2016lqw,Cavaglia:2016oda} was to emphasize how the $T \bar T$ deformation is special and is an exception to this general expectation.

Another interesting feature of $T \bar T$ deformation is that it exhibits Hagedorn behavior in the ultraviolet. An easy way to show this is to first note that a generic undeformed field theory in two dimension will exhibit Cardy behavior in its ultraviolet spectrum\footnote{There are some subtleties in the precise definition of the concept of entropy which we discuss in Appendix \ref{appA}.}
\be S(E) \sim \sqrt{c R E}\ . \ee
Here, we  are taking $c \sim c_L \sim c_R$ to simplify the notation. We will restrict our consideration to set of states with vanishing $P_i$. Then,  for sufficiently large $E$ compared to $R/\mu$, we see from (\ref{def}) that the relation between the deformed and undeformed energy of the $i^{th}$ state is
\be {\cal E}_i = \sqrt{{R E_i \over \mu}} \ . \ee
From this, we can infer that
\be S({\cal E}) \sim  \sqrt{c \mu } {\cal E} \ee
and from the coefficient of the linearly growing term for $S$ as a function of ${\cal E}$, we read off the Hagedorn temperature
\be T^{T \bar T}_H \sim {1 \over \sqrt{c\mu}} \ee
up to numerical constants of order one which are not important for our discussion.\footnote{See \cite{Chakraborty:2020xyz} for a related analysis for the  more general   $\mu T \bar T + \epsilon_+ J \bar T+ \epsilon_- T \bar J$ deformation.}

The fact that the spectrum of generic $T \bar T$ deformed theories is Hagedorn suggests an intimate connection with stringy dynamics and various possible non-local phenomena. Numerous extensions and generalizations have been considered and discussed by many authors which by now are too long to list. We will instead refer to the review article \cite{Jiang:2019epa} to access additional references.

The issue which we wish to pursue in this note is what happens if the undeformed theory has faster than Cardy growth in the density of states. For sake of arguments, let us consider a spectrum which gives rise to the scaling
\be S(E) = \# E^\gamma \label{seg}\ee
for some $\gamma > 1/2$. The same argument presented above leads to the conclusion that if such a spectrum is $T \bar T$ deformed, the resulting entropy for ${\cal E} > R/\mu$ (where the effects of $T \bar T$-deformation dominates) scales according to
\be S({\cal E}) = \#' {\cal E}^{2 \gamma} \ . \label{sgamma} \ee
We see that if $\gamma>1/2$, then the  specific heat
\be {d E \over dT} = {1 \over {d \over d E} \left({dS \over dE}\right)^{-1}} = -T^{-2} \left({d^2 S \over  dE^2} \right)^{-1}=
 T^{-2} \left(\#' (1 - 2 \gamma) 2 \gamma {\cal E}^{2 \gamma - 2}\right)^{-1} \ee
is a negative quantity. Because of this relation, the positivity of specific heat is also expressed as the concavity condition\footnote{There is some subtelty in this statement on which we will elaborate in the discussion that follows.}
\be {d^2 S \over d E^2} < 0 ~.\ee
What this suggests is that if a  conformal field field theory exhibits a spectrum whose entropy grows like (\ref{seg}) for $\gamma > 1/2$, then, it's $T \bar T$ deformation leads to a system which is thermally unstable. 

\section{$T \bar T$ deformed symmetric orbifold CFT}

An immediate question which arises at this point is whether a scaling of the form (\ref{seg}) for $\gamma > 1/2$ ever arises in a conformal field theory. It surely can't arise in the ultraviolet region because one expects a Cardy behavior there. In order for the scaling different than Cardy behavior to be manifested over a large span of energies, some large dimensionless parameter needs to appear as a data in specifying the theory.

\begin{figure}
\centerline{
\begin{tabular}{cc}
\includegraphics[width=3in]{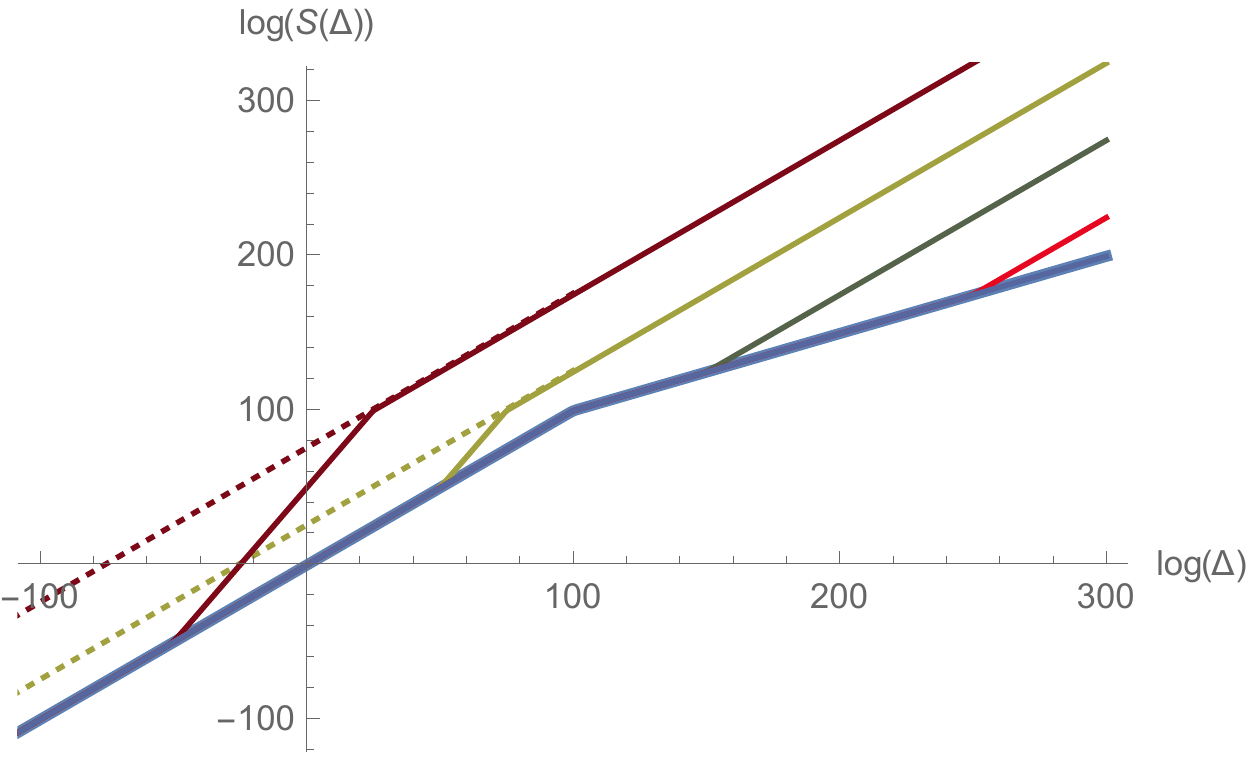} &
\includegraphics[width=3in]{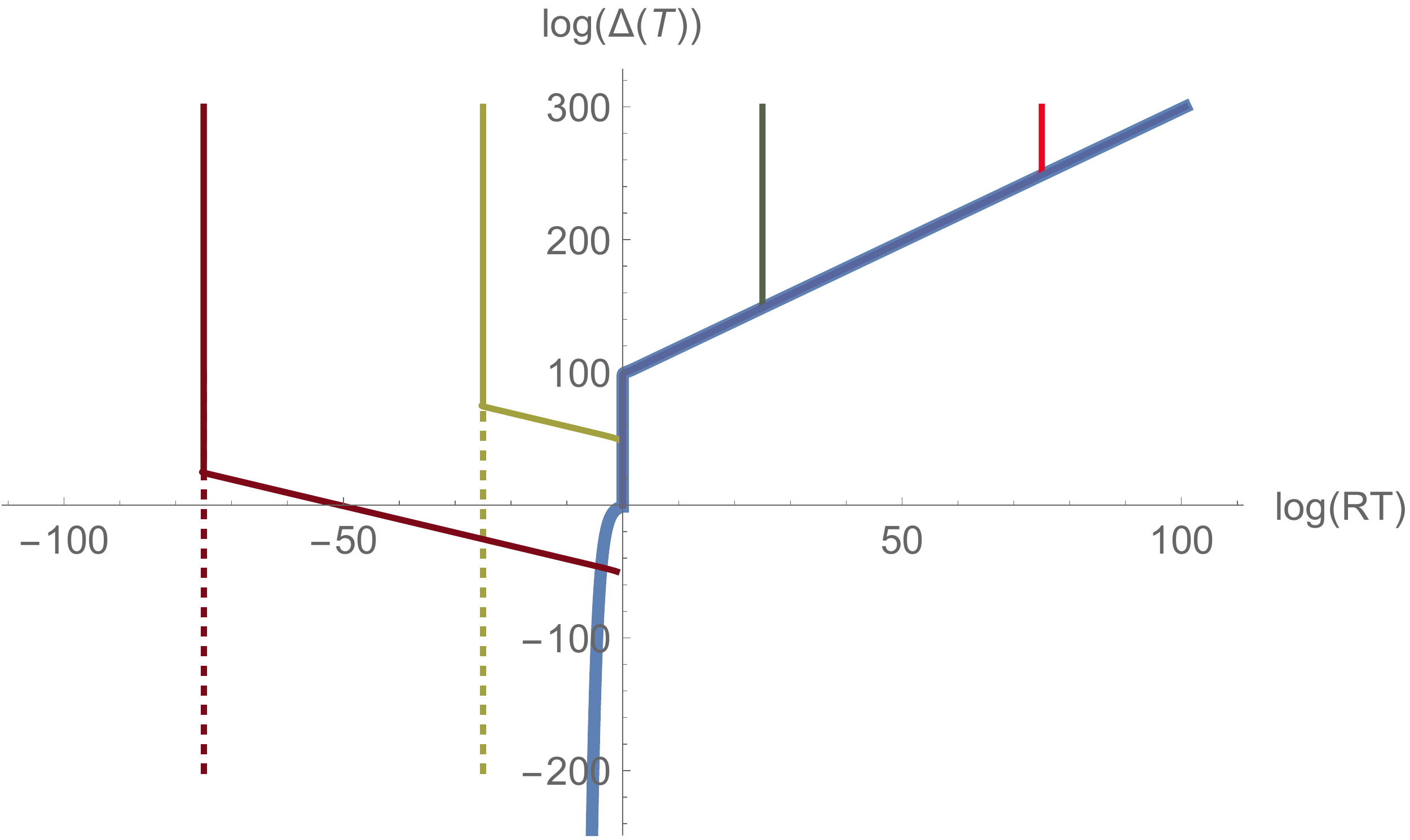} \\
(a) & (b) 
\end{tabular}}
\caption{$S({\cal E})$ and ${\cal E}(T)$ for the $T \bar T$ deformed symmetric product CFT. We actually plot $S(\Delta)$ and $\Delta(T) \equiv R {\cal E}(T) + c_1N/12$ in a log-log plot to schematically highlight the scaling feature. We are using $\Delta(T)$ as a convenient parametrization of energy since it is dimensionless and is zero for the vacuum state.  Where the entropy scales as $S(E) \sim E^\alpha$, the slope of these lines correspond to $\alpha$. When the slope is one, the spectrum has Hagedorn density. Where the slope is $1/2$, the spectrum is Cardy like. The spectrum is super-Hagedorn and exhibits negative specific heat in the segment where the slope is greater than one. The thick blue line describes  the undeformed theory. We have taken $\mu = R^2 (c_1 N)^{-n/2}$, or $T^{T \bar T}_H = (c_1N)^{n/2-1}/R$, for $n$ taking values $-1$, $1$, $3$, and $5$, corresponding to the brown, light green, dark green, and red lines, respectively. The strength of $T \bar T$ deformation increases as one goes from red to brown in the figure. The light green and brown corresponds to the case where  $\mu$, the coefficient of the  $T \bar T$ deformation, is large so that  $T^{T \bar T}_H < T^{HP}\sim 1/R$.    We have taken $c_1N = \exp(100)$ so that the Hagedorn region of the undeformed CFT is hierarchically large.  In the deep IR, $S(\Delta)$ goes from being ${\cal O}(1)$ to zero as $\Delta$ goes from ${\cal O}(1)$ to zero  in accordance with the third law of thermodynamics.   This corresponds to the infrared gap. This will  cause $\Delta(T) \sim  \exp(-1/T)$ to bend slightly to the left  as is illustrated  in plot  (b) and eventually intersect the dotted line at $\Delta \sim \exp\left(-\sqrt{c_1N \mu /R^2}\right)$ as illustrated in figure \ref{figb}.  The dotted line indicates the expected behavior of $\Delta(T)$ for the system coupled with a heat bath through Gibbsian ruling. The case illustrated with the brown curve ($n=-1$)  has the $T \bar T$ deformation parameter $\mu$ set so large that the onset of $T_H^{T \bar T}$ Hagedorn spectrum is below the scale of the gap, but these features are shielded by the Gibbsian ruling. Strictly speaking, the first order phase transition temperature illustrated with the dotted line is slightly smaller than the $T_H$. This is hard to see in a log-log plot but is illustrated better in figure \ref{figb} which is a linear plot.  What appears as sharp bends are actually smooth transitions for large but finite $N$. Some numerical factors of order one have been massaged to make the figure fit without changing the general feature. As such these figures should be considered schematic.
\label{figa}}
\end{figure}

Fortunately, one does not need to look very hard to find such a construction. The large $N$ symmetric product CFT
\be {\cal M}_N = {\cal M}^N/S_N~, \, \ee
where ${\cal M}$ represents some CFT with central charge $c_1$,
exhibits such behavior, as was analyzed in \cite{Keller:2011xi}. We use the notation $c_1$ to avoid confusion between the central charge of ${\cal M}$ and the central charge $c_1 N$ of ${\cal M}^N/S_N$. (We will also be  considering compact CFTs ${\cal M}$ whose spectrum is discrete.) The analysis in \cite{Keller:2011xi} was mostly in the canonical ensemble, but the corresponding microcanonical expression can be found in  (1.12) and (5.8) of \cite{Hartman:2014oaa}. Let us take $c_1$ to be of order one\footnote{We will take $N$ to be a large but finite integer. We will also explore wide range of values for $N^{-5/2}< \mu/R^2 <  N^{1/2}$ as is illustrated  in figure \ref{figa}. $R$ sets the over-all scale of the problem.} as opposed to scaling with $N$. One can infer the scaling behavior of the entropy for the set of states with vanishing  momentum $P$ as\footnote{In \cite{Hartman:2014oaa}, entropy for fixed potential conjugate to $P$ was computed, but one expects fixed $P=0$ entropy and the fixed $\mu=0$ entropy to be the same in the thermodynamic limit. One way to see this is to note that $e^{4 \pi \sqrt{m \bar m}} \sim e^{2 \pi \sqrt{E^2-P^2}}$ in (2.6) of \cite{Keller:2011xi} is peaked at $P=0$. In considering the $T \bar T$ deformation, the sum over $P$ at most contributes terms depending logarithmically on ${\cal E}$ to $S({\cal E})$.}
\be S(E) \sim \left\{ \begin{array}{ll}
2 \pi {\sqrt{c_1N  \Delta} \over 3} & \text{ for }\ {c_1N \over 6 } \ll \Delta \\
2 \pi \Delta  & \text{ for }\ {{\cal O}(1)  \ll \Delta \ll  {c_1N  \over 6 }} \\
{\cal O}(1) \rightarrow 0 & \text{ for }\  0  < \Delta < {\cal O}(1)\end{array} \right. , \label{SofE} \ee
where
\be \Delta = RE + {c_1 N \over 12} \ee
is the dimension of the operators of the symmetric orbifold CFT associated with the state of energy $E$.  The minimum energy $E_{min} =  -{c_1 N / 12 R}$ is the vacuum Casimir energy.  
The size of the intermediate Hagedorn scaling region increases for large $N$. Above that region,  in the ultraviolet, the spectrum is that of Cardy.\footnote{The entropy $S(\Delta)$ (\ref{SofE}) matches at the cross-over points  $\Delta \sim c_1N/6$ and $\Delta \sim {\cal O}(1)$ in the spirit of \cite{Itzhaki:1998dd}.} This Hagedorn behavior can  be interpreted as corresponding to a first order phase transition which is in the  universality class as  Hawking-Page phase transition \cite{Witten:1998zw} which is expected to take place at the temperature of order
\be T^{HP} \sim {1 \over R} \  \ee
for CFT's which admit a gravity dual.  We expect the spectrum to be dense down to $\Delta  \sim {\cal O}(N^0) \equiv{\cal O}(1)  $ corresponding to the operator of lowest dimension other than the identity operator in the symmetric orbifold theory. These low lying states comes from the untwisted sector of the symmetric orbifold CFT\footnote{We thank T.~Hartman for discussion on this point.} \cite{Klemm:1990df}. Strictly speaking, the system is gapped in this region and the thermodynamic approximation breaks down. We will only use the fact that  that $S(T) \rightarrow 0$ and $\langle E \rangle \rightarrow -c_1N/12R$ as $T \rightarrow 0$ to be consistent with the third law of thermodynamics (and the existence of a non-degenerate vacuum) which we represent as ${\cal O}(1) \rightarrow 0$ in equation (\ref{SofE}).

The effect of $T \bar T$ deformation on this spectrum can be obtained by substituting
\be E = {\cal E}+ {\mu \over R} {\cal E}^2 \ee
into $S(E)$ given in (\ref{SofE}) illustrated in figure  \ref{figa}.

The behavior illustrated in figure \ref{figa} is quite reasonable. If $T^{T \bar T}_H > T^{HP}$, then one finds a cross-over from the Cardy behavior to the $T \bar T$ deformed Hagedorn behavior at ${\cal E} \sim R/\mu$ where  $T =T^{T \bar T}_H=\sqrt{1 / c_1 N \mu}$. Things gets a bit more interesting when $T^{T \bar T}_H < T^{HP}$. In this case, the density of states must interpolate between $S({\cal E}) = {\cal E}/T^{HP}$ behavior at ${\cal E} < R/ \mu$, then cross over to the interpolating behavior $S({\cal E}) \sim \mu {\cal E}^2$ in the range $R/\mu < {\cal E} < \sqrt{c_1N/\mu}$,  and $S({\cal E}) = {\cal E}/T^{T \bar T}_H$ for $ \sqrt{c_1N/\mu} < {\cal E}$. Note that in the interpolating region, the entropy scales as ${\cal E}^2$.

The take away message at this point is that the specific heat associated with this system, in the interpolating region, is negative. Thermodynamics of systems with negative specific heat is a subject which by now is well established \cite{thirring,fisher,Lynden-Bell:1998pzh}, but some aspects can be subtle and confusing at a first pass. We will therefore make several comments about the negative specific heat using the $T \bar T$ deformed symmetric product CFT as a concrete example to highlight some of the basic points. 

\section{Negative Specific Heat and the Gibbsian Ruling}

One standard lore is that given a partition function $Z(\beta)$, one can write
\be T^2 {d \over d T} \langle E \rangle = {d \over d \beta^2} \log (Z(\beta)) = {Z''(\beta) \over Z(\beta) } - \left({Z'(\beta) \over Z(\beta)} \right)^2 = \langle E^2 \rangle - \langle E \rangle^2 = \Delta E^2 > 0 ~,\ee
which appears to suggest that if a quantum system has a spectrum for which the Boltzmann partition sum is well defined, the specific heat $d \langle E \rangle / d T$ is always positive. This appears then to  imply that
\be {\partial^2 S \over d E^2} = {d \over dE} {1 \over T} = -T^2 {dT \over dE} < 0 \label{bound}\ee
for $S(E)$ in \eqref{sgamma} with $\gamma>1/2$, but there is a subtlety which invalidates this statement.\footnote{In \cite{thirring}, the hydrogen-like spectrum is presented as a concrete example of a well defined spectrum giving rise to a negative specific heat, violating the bound (\ref{bound}).} What is correct is that $\langle E \rangle(T)$ is a monotonic function in $T$ provided that the system is coupled to a heat bath and that the thermodynamic equilibrium is achieved. When the specific heat is negative, however, $S(T)$ is not a single valued function. One can still define a Boltzmann sum over these states and $\langle E \rangle (T)$ will be monotonic, but the system undergoes a phase transition jumping from one stable branch to another satisfying the condition analogous to the Clausius–Clapeyron condition.\footnote{It is straight forward to carry out an exercise where one assumes a concrete form of $S(E)$ which leads to a non-single valued $S(T)$ and then to explicitly compute $\langle E \rangle(T)$ for a Boltzman distribution which necessarily comes out as a single valued function.} This mechanism is referred to as the ``Gibbsian Ruling'' in \cite{fisher}.  

\begin{figure}
\begin{tabular}{cc}
\includegraphics[width=3in]{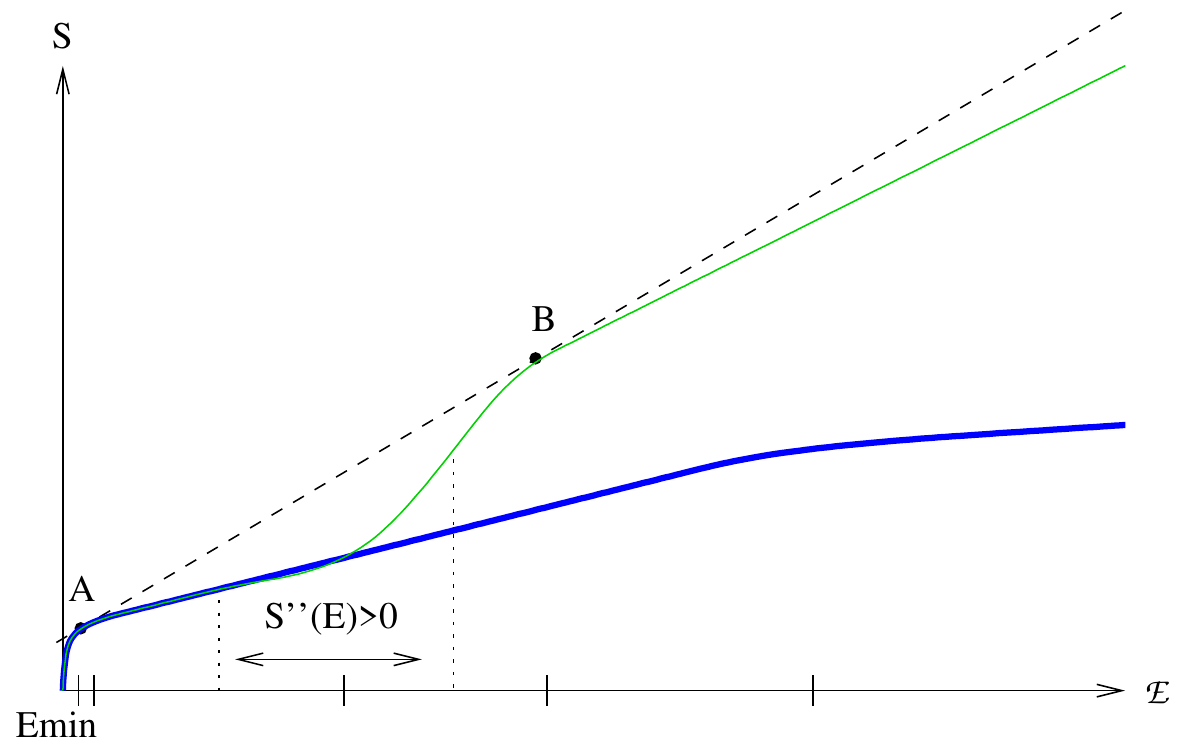} &
\includegraphics[width=3in]{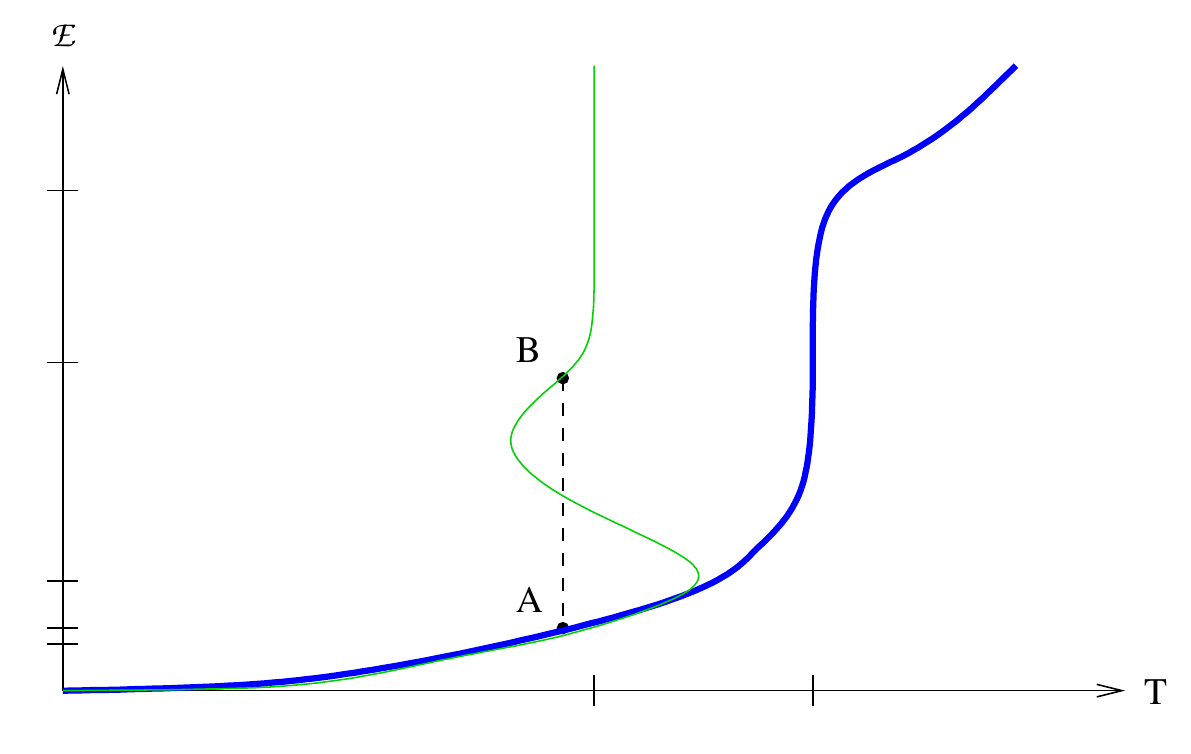} \\
(a)&(b) \end{tabular}
\caption{(a) $S({\cal E})$ when $1/c_1N \ll   \mu/R^2 \ll 1$ in linear plot is illustrated schematically in green. The same data was also illustrated in green in figure \ref{figa}.a in a log-log plot. In order to fit all of the features in one plot, this figure is not drawn to scale. For large $E$, the entropy $S({\cal E})$ asymptotes to grow linear in ${\cal E}$  due to the Hagedorn spectrum of the $T \bar T$ deformation. The dotted line between the points {\bf A} and {\bf B} indicates the effect of Gibbsian ruling which eliminates the region which contians the interval with negative specific heat (where $S''({\cal E})>0$) and gives rise to a first order phase transition.  The slope of the $S({\cal E})$ curve corresponds to $T^{-1}$. This plot shows that the temperature associated with the Gibbsian ruling illustrated using the dotted line is slightly smaller than the temperature associated with the Hagedorn temperature of the $T \bar T$ deformation. The dotted line is tangent to the $S({\cal E})$ curve at points {\bf A} and {\bf B}. Point {\bf B} appears in figure \ref{figa}.a. Point {\bf A} appears as a result of almost parallel lines meeting at $\Delta \sim {\cal O}(1)$ because of ${\cal O}(1)$ corrections which are hard to see in the logarithmic scale used in figure \ref{figa}.a. The five tick marks on the horizontal axis corresponds to ${\cal E}-E_{min}$ of order $\exp(-\sqrt{c_1N \mu/R^2})/R$, $1/R$, $R/\mu$,  $\sqrt{c_1N/\mu}$, and $c_1N/R$, respectively in the increasing order. It should be stressed that ${\bf B}$ occurs near ${\cal E} \sim \sqrt{c_1N/\mu}$ which is set hierarchically larger than ${\bf A}$ which is at ${\cal E} - E_{min} \sim \exp\left(-\sqrt{c_1N \mu/R^2}\right)$. The blue line is $S(E)$ of the symmetric orbifold theory without any $T \bar T$-deformation. At $E \sim c_1 N/R$ (the right most tick mark), the blue curve is crossing over from the Hagedorn behavior $S \sim RE$ to Cardy behavior $S \sim \sqrt{c_1 N R E}.$ (b) ${\cal E}(T)$ for the same system. The five tick marks on the ${\cal E}$ axis are the same as plot (a). The tick marks on the $T$ axis corresponds to $T = T_H^{T \bar T} \sim 1/\sqrt{c_1 N \mu}$ and $T=T^{HP}\sim1/R$.
\label{figb}}
\end{figure}

In other words, when a system is isolated, one can characterize its thermodynamic properties using the microcanonical parameters, and one can explore the regions in parameter space where the specific heat is negative. As soon as a heat bath is introduced, however, the unstable states are physically driven toward the coexistence state of stable states. Negative specific heat of the $T \bar T$ deformed symmetric product CFT simply indicate that the system undergoes a first order phase transition. Since the symmetric product CFT exhibited Hawking-Page phase transition even prior to the $T \bar T$ deformation, all this means is that the Hagedorn behavior of the $T \bar T$ deformation and that of Hawking-Page merge in the phase diagram. In  more concrete terms, one expects the Hagedorn branch illustrated by the vertical dotted lines in \ref{figa}.b to continue until it meets the thick blue line bending toward small values of $T$ as $\Delta$ is decreased.  In \ref{figa}.a, the Hagedorn line of slope one simply continues toward small $\Delta$ but its slope gradually decreases so that $T$ approaches zero as $\Delta$ approaches zero.  This is best seen by drawing the $S({\cal E})$ in linear, rather than in logarithmic, scale as is illustrated in figure \ref{figb}.a.

To reiterate, spectrum giving rise to negative specific heat in microcanonical ensemble gives rise to first order phase transition in the canonical ensemble.\footnote{Similar points were made about black holes in  asymptotically anti de-Sitter space in \cite{Chamblin:1999tk,Cvetic:1999ne}.} Our claim then is simply the statement that $T \bar T$ deformation of large $N$ symmetric product CFT gives rise to such a system. It is important to keep in mind, however, that the system cannot be in equilibrium at temperature higher than $T^{T\bar T}_H$ (when $T^{HP}<T^{T\bar T}_H$) since the UV spectrum is Hagedorn due to $T \bar T$ deformation. When $T_H^{T \bar T} < T^{HP}$, $T_H^{T \bar T}$ is the critical temperature for the phase transition. 

Similar general discussion about negative specific heat in $T \bar T$ deformed conformal field theories and in quantum mechanics was discussed\footnote{The ``Gibbsian ruling'' in \cite{fisher} is referred to as ``Maxwell construction'' in \cite{Barbon:2020amo}.} in \cite{Barbon:2020amo}. The physical interpretation that the negative specific heat gives rise to first order phase transition in canonical ensemble is explained there as well. In \cite{Barbon:2020amo}, the possibility of negative specific heat was explored as  arising from the small correction to the Hagedorn behavior of the $T \bar T$ deformation of a system with Cardy like density of states. What we provide here is a simple and concrete realization of negative specific heat in the $T \bar T$ deformed large $N$ symmetric product CFT system. 

In figure \ref{figb}, we illustrate $S(E)$ for our system schematically so that the Gibbsian ruling can be seen more easily than in figure \ref{figa}.a.

\section{Discussions}

Now that we understand that there is a first order phase transition for $T^{T\bar{T}}_H<T^{HP}$, in the $T \bar T$-deformed large $N$ symmetric orbifold CFT, there are a number of interesting questions one might explore. One is whether there is an order parameter which can be used to distinguish the different phases at the critical temperature. It is natural to expect that some CFT operator could function as an order parameter. Perhaps it is a Wilson loop type operator as was discussed in \cite{Witten:1998zw}. Once the order parameter is identified, it would be very interesting to look for a domain wall configuration interpolating between the two phases. This issue can also be explored for the Hawking-Page transition itself, and should admit a concrete realization as a solution on the gravity side of the holographic correspondence. 

In this article, we  considered the double trace $(\sum T_i)(\sum \bar T_i)$ deformation, where $T_i$ refers to the stress energy tensor of the $i^{th}$ block of the product CFT. In a symmetric product CFT, one can also consider the single trace $\sum_i T_i \bar T_i$ deformation which has interesting features especially on the gravity side of the holographic correspondence \cite{Giveon:2017nie,Hashimoto:2019hqo,Chakraborty:2020swe}.  It would be interesting to explore the thermodynamic equation of states for the single trace $T\bar T$ deformed symmetric product CFT  and explore the interplay between Hawking Page and $T\bar{T}$ scales \cite{Chakraborty:2018aji,Chakraborty:2020xyz}.

Another interesting question is whether there are other situations where $1+1$ field theory exhibits scaling $\gamma > 1/2$ for the density of states (\ref{seg}).  One simple system exhibiting such a feature is the large $N$ supersymmetric Yang-Mills theory in 1+1 dimensions with 16 supercharges \cite{Itzhaki:1998dd}. This model contains states whose density scales as
\be S(E) \sim   \sqrt{N} g_{YM}^{-1/3} E^{2/3} R^{1/3}, \qquad {g_{YM} \over \sqrt{N}}< E < g_{YM} \sqrt{N} .\ee
Naively, we have
\be \gamma = {2 \over 3} > {1 \over 2} \ . \ee
One must however account for the fact that the 1+1 SYM is not a conformal field theory, and as such, we can not blindly apply (\ref{def}).  Nonetheless, one expects from experience that the effects of $T \bar T$ deformation tends to be universal. It seems reasonably likely that the $T \bar T$ deformed 1+1 SYM will also exhibit first order phase transition in the canonical ensemble.

\section*{Acknowledgements}

We would like to thank W. Cottrell, A. Dymarsky, S. Georgescu, M. Guica, T. Hartman and D. Kutasov for helpful discussions. The work of SC is supported in part by the ERC Starting Grant 679278 Emergent-BH.

\appendix

\section{Microcanonical and thermal entropy \label{appA}}

Entropy $S(E)$ is defined in terms of the spectrum by the relation
\be S_{micro}(E) = \log (\overline{\rho(E)})  = (\log \overline{\sum_i \delta(E - E_i)}) , \ee
where the overline refers to smearing over some range of energies
\be \overline {\rho(E)} = \int d \epsilon \, f(L \epsilon) \rho(E+\epsilon), \ee
where $f(L \epsilon)$ is some smearing function centred around $\epsilon=0$, and $L$ is a constant with dimension of length. There is some arbitrariness in the definition of $S_{micro}(E)$ which follows from the arbitrariness of $f(L \epsilon)$ but this arbitrariness is sub-leading in the thermodynamic (large central charge) limit.  More discussions on this subtle issue can be found in appendix A of \cite{Hartman:2014oaa}. We can also define $S(E)$ parametrically in terms of
\be S_{thermo} (T) = {d \over d T} T \log Z(T), \qquad \langle E \rangle = T^2 {d \over dT} \log Z(T) , \qquad Z(T) = \int_{E_{min}}^\infty  dE\, \rho(E) e^{-{E \over T}}\ee
and we expect $S_{micro}(E)$ and $S_{thermo}(\langle E \rangle)$ to agree in the thermodynamic limit as long as the equation of state $S(T)$ is single valued.  We will henceforth drop the subscript ``micro'' and ``thermo.'' Third law of thermodynamics will imply that $S(E)$ approaches a constant $S_0$ as $E \rightarrow 0$. We can set that constant  $S_0 = S(E=E_{min})=0$ by adjusting the overall normalization of $f(L \epsilon)$. $S_{thermo}(\langle E \rangle)$ automatically goes to zero and $\langle E \rangle  = E_{min}$  when $T=0$.

\providecommand{\href}[2]{#2}\begingroup\raggedright\endgroup


\begin{thebibliography}{10}

\bibitem{Smirnov:2016lqw}
F.~A. Smirnov and A.~B. Zamolodchikov, ``{On space of integrable quantum field
  theories},'' {\em Nucl. Phys.} {\bf B915} (2017) 363--383,
\href{http://www.arXiv.org/abs/1608.05499}{{\tt 1608.05499}}.

\bibitem{Cavaglia:2016oda}
A.~Cavaglià, S.~Negro, I.~M. Szécsényi, and R.~Tateo, ``{$T
  \bar{T}$-deformed 2D Quantum Field Theories},'' {\em JHEP} {\bf 10} (2016)
  112,
\href{http://www.arXiv.org/abs/1608.05534}{{\tt 1608.05534}}.

\bibitem{McGough:2016lol}
L.~McGough, M.~Mezei, and H.~Verlinde, ``{Moving the CFT into the bulk with $
  T\overline{T} $},'' {\em JHEP} {\bf 04} (2018) 010,
  \href{http://www.arXiv.org/abs/1611.03470}{{\tt 1611.03470}}.

\bibitem{Chakraborty:2020xyz}
S.~Chakraborty and A.~Hashimoto, ``{Thermodynamics of $
  \mathrm{T}\overline{\mathrm{T}} $, $ \mathrm{J}\overline{\mathrm{T}} $, $
  \mathrm{T}\overline{\mathrm{J}} $ deformed conformal field theories},'' {\em
  JHEP} {\bf 07} (2020) 188, \href{http://www.arXiv.org/abs/2006.10271}{{\tt
  2006.10271}}.

\bibitem{Jiang:2019epa}
Y.~Jiang, ``{A pedagogical review on solvable irrelevant deformations of 2D
  quantum field theory},'' {\em Commun. Theor. Phys.} {\bf 73} (2021), no.~5,
  057201, \href{http://www.arXiv.org/abs/1904.13376}{{\tt 1904.13376}}.

\bibitem{Keller:2011xi}
C.~A. Keller, ``{Phase transitions in symmetric orbifold CFTs and
  universality},'' {\em JHEP} {\bf 03} (2011) 114,
  \href{http://www.arXiv.org/abs/1101.4937}{{\tt 1101.4937}}.

\bibitem{Hartman:2014oaa}
T.~Hartman, C.~A. Keller, and B.~Stoica, ``{Universal Spectrum of 2d Conformal
  Field Theory in the Large c Limit},'' {\em JHEP} {\bf 09} (2014) 118,
  \href{http://www.arXiv.org/abs/1405.5137}{{\tt 1405.5137}}.

\bibitem{Itzhaki:1998dd}
N.~Itzhaki, J.~M. Maldacena, J.~Sonnenschein, and S.~Yankielowicz,
  ``{Supergravity and the large N limit of theories with sixteen
  supercharges},'' {\em Phys. Rev. D} {\bf 58} (1998) 046004,
  \href{http://www.arXiv.org/abs/hep-th/9802042}{{\tt hep-th/9802042}}.

\bibitem{Witten:1998zw}
E.~Witten, ``{Anti-de Sitter space, thermal phase transition, and confinement
  in gauge theories},'' {\em Adv. Theor. Math. Phys.} {\bf 2} (1998) 505--532,
  \href{http://www.arXiv.org/abs/hep-th/9803131}{{\tt hep-th/9803131}}.

\bibitem{Klemm:1990df}
A.~Klemm and M.~G. Schmidt, ``{Orbifolds by Cyclic Permutations of Tensor
  Product Conformal Field Theories},'' {\em Phys. Lett. B} {\bf 245} (1990)
  53--58.

\bibitem{thirring}
W.~Thirring, ``{Phases and phase diagrams: Gibbs' legacy today},'' {\em Z.
  Physik} {\bf 235} (1970) 339–352.

\bibitem{fisher}
M.~Fisher, ``{Systems with negative specific heat},'' {\em Proceedings of the
  Gibbs Symposium} (1989) 39--72.

\bibitem{Lynden-Bell:1998pzh}
D.~Lynden-Bell, ``{Negative specific heat in astronomy, physics and
  chemistry},'' {\em Physica A} {\bf 263} (1999) 293--304,
  \href{http://www.arXiv.org/abs/cond-mat/9812172}{{\tt cond-mat/9812172}}.

\bibitem{Chamblin:1999tk}
A.~Chamblin, R.~Emparan, C.~V. Johnson, and R.~C. Myers, ``{Charged AdS black
  holes and catastrophic holography},'' {\em Phys. Rev. D} {\bf 60} (1999)
  064018, \href{http://www.arXiv.org/abs/hep-th/9902170}{{\tt hep-th/9902170}}.

\bibitem{Cvetic:1999ne}
M.~Cvetic and S.~S. Gubser, ``{Phases of R charged black holes, spinning branes
  and strongly coupled gauge theories},'' {\em JHEP} {\bf 04} (1999) 024,
  \href{http://www.arXiv.org/abs/hep-th/9902195}{{\tt hep-th/9902195}}.

\bibitem{Barbon:2020amo}
J.~L.~F. Barbon and E.~Rabinovici, ``{Remarks on the thermodynamic stability of
  $T \bar T$ deformations},'' {\em J. Phys. A} {\bf 53} (2020), no.~42, 424001,
  \href{http://www.arXiv.org/abs/2004.10138}{{\tt 2004.10138}}.

\bibitem{Giveon:2017nie}
A.~Giveon, N.~Itzhaki, and D.~Kutasov, ``{$ \mathrm{T}\overline{\mathrm{T}} $
  and LST},'' {\em JHEP} {\bf 07} (2017) 122,
  \href{http://www.arXiv.org/abs/1701.05576}{{\tt 1701.05576}}.

\bibitem{Hashimoto:2019hqo}
A.~Hashimoto and D.~Kutasov, ``{Strings, symmetric products, $T \bar{T}$
  deformations and Hecke operators},'' {\em Phys. Lett. B} {\bf 806} (2020)
  135479, \href{http://www.arXiv.org/abs/1909.11118}{{\tt 1909.11118}}.

\bibitem{Chakraborty:2020swe}
S.~Chakraborty, A.~Giveon, and D.~Kutasov, ``{$ T\overline{T} $, black holes
  and negative strings},'' {\em JHEP} {\bf 09} (2020) 057,
  \href{http://www.arXiv.org/abs/2006.13249}{{\tt 2006.13249}}.

\bibitem{Chakraborty:2018aji}
S.~Chakraborty, ``{Wilson loop in a $T\bar{T}$ like deformed $\rm{CFT}_2$},''
  {\em Nucl. Phys. B} {\bf 938} (2019) 605--620,
  \href{http://www.arXiv.org/abs/1809.01915}{{\tt 1809.01915}}.

\end{thebibliography}
\end{document}